\documentclass[fleqn,usenatbib]{mnras}

\usepackage{newtxtext} 
\usepackage{mathptmx}
\usepackage{txfonts}
\usepackage{bm} 
\usepackage[T1]{fontenc}
\usepackage{xcolor} 
\DeclareRobustCommand{\VAN}[3]{#2}
\let\VANthebibliography\thebibliography
\def\thebibliography{\DeclareRobustCommand{\VAN}[3]{##3}\VANthebibliography}

\usepackage{graphicx}	
\usepackage{amssymb}	

\DeclareRobustCommand{\ion}[2]{%
\relax\ifmmode
\ifx\testbx\f@series
{\mathbf{#1\,\mathsc{#2}}}\else
{\mathrm{#1\,\mathsc{#2}}}\fi
\else\textup{#1\,{\mdseries\textsc{#2}}}%
\fi}
\title[Strength in Numbers: Red Galaxies Bolster the Cosmic Star Formation Rate Density at  $z \gtrsim 3$] {Strength in Numbers: Red Galaxies Bolster the Cosmic Star Formation Rate Density at  $z \gtrsim 3$}
\author[L. Barrufet et al.]{L. Barrufet$^{1}$\thanks{E-mail:lbarrufe@edu.ac.uk}, J.S. Dunlop$^{1}$, R. Begley$^{1}$,
S. Flury$^{1}$,
D.J. McLeod$^{1}$,
K. Arellano-Cordova$^{1}$,
A. Carnall$^{1}$, 
F. Cullen$^{1}$, \newauthor
C. T. Donnan$^{2}$, 
F. Liu$^{1}$, 
R. McLure$^{1}$, 
D. Scholte$^{1}$,
T. M. Stanton$^{1}$, 
R. Cochrane$^{3}$,
C. Conselice$^{3}$, \newauthor
R. Ellis $^{4}$, 
P. G. P\'erez-Gonz\'alez$^{5}$, 
R. Gottumukkala$^{6, 7}$, 
N. A. Grogin$^{8}$, 
G. D. Illingworth$^{9}$,
A. M. Koekemoer$^{8}$, \newauthor
D. Magee$^{9}$, 
M. Michalowski$^{10}$ \\ 
$^{1}$ Institute for Astronomy, University of Edinburgh, Royal Observatory, Edinburgh EH9 3HJ, UK \\
$^{2}$ NSF's National Optical-Infrared Astronomy Research Laboratory, 950 N. Cherry Ave., Tucson, AZ 85719, USA\\
$^{3}$ Jodrell Bank Centre for Astrophysics, University of Manchester, Manchester M13 9PL, UK \\
$^{4}$ Department of Physics \& Astronomy, University College London, Gower St., London WC1E 6BT, UK \\
$^{5}$ Centro de Astrobiolog\'{\i}a (CAB), CSIC–INTA, Cra. de Ajalvir km 4, 28850- Torrej\'on de Ardoz, Madrid, Spain \\
$^{6}$ Cosmic Dawn Center (DAWN), Niels Bohr Institute, University of Copenhagen, Jagtvej 128, København N, DK-2200, Denmark \\
$^{7}$ Niels Bohr Institute, University of Copenhagen, Jagtvej 128, Copenhagen, Denmark \\
$^{8}$ Space Telescope Science Institute, 3700 San Martin Drive, Baltimore, MD 21218 \\
$^{9}$ Department of Astronomy and Astrophysics, UCO/Lick Observatory, University of California, Santa Cruz, CA 95064, US \\
$^{10}$ Astronomical Observatory Institute, Faculty of Physics and Astronomy, Adam Mickiewicz University, ul. Słoneczna 36, 60-286 Poznan, Poland \\
}

\date{Submitted}

\pubyear{2025}

\begin{document}
\label{firstpage}
\pagerange{\pageref{firstpage}--\pageref{lastpage}}
\maketitle

\begin{abstract} 
A comprehensive account of the cosmic star-formation history demands an accurate census of dust-enshrouded star formation over cosmic time. We provide strong new constraints from a large sample of 777 red galaxies, selected based on their dust-reddened, rest-frame UV-optical emission. This sample of 777 galaxies spans $1 < z < 8$ and is selected from PRIMER JWST NIRCam and HST COSMOS optical data, ensuring robust colour criteria. The SEDs indicate that these dust-reddened galaxies are star-forming, with median $\mathrm{SFR \sim 40M_{\odot}yr^{-1}}$ and stellar mass $\log(M_{*}/M_{\odot}) = 10.3^{+0.6}_{-0.8}$; each exceeds the corresponding medians of the full JWST-detected population by over two dex. Our sample thus clearly shows that red galaxies dominate the high-mass end: they comprise 72 \% of galaxies with $\log(M/M_{\odot}) > 10$ at $z = 3.3$, rising to 91\% by $z \sim 7$ (albeit with large uncertainties at the highest redshifts). Crucially, we find that the number density of massive red star-forming galaxies at $z \sim 6$ is sufficient to explain the abundance of quiescent galaxies at $z > 3$, consistent with typical quenching timescales allowed in the $\mathrm{\sim 1Gyr}$ interval from $z \sim 6$ to $z \sim 3$. This large abundance yields a substantial contribution to the cosmic star-formation rate density: at $z \sim 4$, red galaxies provide $\mathrm {\rho_{SFR} = 3.9^{+0.6}_{-0.5} \times 10^{-2} M_{\odot} yr^{-1}Mpc^{-3}}$, and at $z \sim 5$ they supply nearly 40 \% of the total $\rho_{SFR}$. This exceeds the contribution of bright sub(mm)-selected dusty star-forming galaxies by more than an order of magnitude. Future deeper and wider ALMA surveys will provide further opportunities to strengthen and extend our results in our quest to fully quantify the contribution of dust-obscured activity to $\rho_{\mathrm{SFR}}$ at high redshifts.

\end{abstract}


\section{Introduction}

Over the past two decades, our understanding of galaxy evolution at redshifts $z>3$ has been predominantly shaped by the study of rest-frame ultraviolet-selected galaxies, due in part to the massive impact of the {\it Hubble Space Telescope} ({\it HST}) at optical wavelengths, but also because of the limited availability of deep infrared--to--mm observations. Nonetheless, the existence and potential importance of very red galaxies have been known for some time (e.g., \citealt{Smail2002, Caputi2004, Dunlop2007, Huang2011}), and subsequent {\it Spitzer} IRAC and Atacama Large Millimeter Array (ALMA) studies have indicated that optically-faint galaxies might be more numerous than perhaps anticipated (e.g., \citealt{Caputi2012, Caputi2015, Wang2016, Franco2018, Wang2019, AlcaldePampliega2019}). 

However, there is no doubt that both the relative lack of deep near/mid-infrared imaging and spectroscopic data at wavelengths $\lambda > 1.6\,{\rm \mu m}$ before the {\it James Webb Space Telescope} ({\it JWST}), as well as the very limited sky coverage achieved to date with ALMA, have hindered comprehensive and complete studies of these red, likely dusty objects. As a result, the contribution of dust-obscured star formation to the cosmic star-formation rate density 
(SFRD = $\rho_{\rm SFR}$), while now fairly well established at `cosmic noon' ($z \simeq 2$; \citealt{Dunlop2017,Michalowski2017}), it has until now remained highly uncertain at higher redshifts
\citep{Casey2018, Casey2021, Wang2019a, Gruppioni2020, Khusanova2021, Talia2021, Barrufet2023b, Zavala2021}.

Now, the unprecedented infrared imaging (both in terms of depth and extended wavelength coverage) and spectroscopy provided by {\it JWST} are revolutionising this field. Specifically, early imaging observations with {\it JWST} have already revealed many red galaxies at high redshifts that were previously missed in earlier surveys \citep[e.g.,][]{Barrufet2023, PerezGonzalez2023, Nelson2022, Rodighiero2023, Labbe2023a, GomezGuijarro2023, Gottumukkala2024, Williams2024}. Despite variations in colour selection techniques and terminology (e.g., {\it HST}-dark, optically-dark, $H$-dropouts), these studies have consistently identified red/optically-faint galaxies at redshifts $\mathrm{z \sim 3-8}$. While red colour selections can include (likely rare) quiescent galaxies and even $\mathrm{z > 10}$ sources \citep{Rodighiero2023}, there is broad agreement that most of the galaxies unveiled by these selection techniques are heavily dust-obscured. In particular, photometric studies agree that dust attenuation can reach values as high as $\mathrm{A_{v} \sim 6}$, with a median of $\mathrm{A_v = 1.9 \pm 0.4}$ \citep{Dunlop2007, Barrufet2023, Rodighiero2023, GomezGuijarro2023}. Most recently, \citet{Barrufet2024} used {\it JWST} NIRSpec data to quantify that $\simeq 90$\% of red sources at $z>3$ are dusty, with the Balmer decrement in these red sources indicating $> 2$\,mag. of attenuation.

Despite this swift progress, the physical properties, including in particular the stellar masses of red galaxies, remain a topic of scrutiny and debate. While red colour selections in the $\lambda_{\rm obs} = 1.5 - 4.4\,{\rm \mu m}$ range are commonly used across many studies, some also impose a selection cut based on rest-frame optical detection, which can bias stellar mass estimates. The {\it JWST} NIRSpec study recently conducted by \citet{Barrufet2024} avoided the imposition of such additional cuts and, aided by robust spectroscopic redshifts, found that red galaxies have uniformly high stellar masses, typically $\log(M_{*}/{\rm M_{\odot}}) > 9.8$. More broadly, the claimed discovery of high-mass galaxies at very high redshifts has raised the question of whether such extreme objects challenge current galaxy formation models or even our understanding of cosmology \citep{Labbe2023a}. However, there is growing evidence that some of the masses of such objects may have been overestimated due to the presence of an Active Galactic Nucleus (AGN), and the inclusion of longer-wavelength data has been found to temper some of the more extreme derived stellar-mass values (due to improved constraints on stellar population age, etc;  \citealt{Wang2024, Williams2024}). 
Nevertheless, the basic finding that most red galaxies are much more massive than galaxies in general appears to be secure, and indeed \citet{Gottumukkala2023} have demonstrated that dusty galaxies likely dominate the high-mass end of the stellar mass function at high redshifts, with $\simeq 20$\% of the most massive galaxies having been missed by {\it HST} at  $3 < z < 6$, rising to nearly 45\% at $6 < z < 8$ (a finding confirmed by \citealt{Weibel2024}). 

Recently, to further complicate the picture, a new population of very compact, red sources, termed `little red dots' (LRDs), has been uncovered through deep {\it JWST} NIRCam imaging surveys \citep{Labbe2023a, Matthee2023b}. These sources are generally characterised by a V-shaped spectral energy distribution (SED), have a point-like appearance (at least at {\it JWST} resolution) and appear to be confined to redshifts $z > 4$ \citep{Kocevski2024}. Despite now being the subject of several imaging studies \citep{Akins2023, Greene2023, Gentile2024LRD} and spectroscopic investigations \citep{Kokorev2023, Matthee2023b}, the question of whether the light from LRDs is dominated by dust-obscured black-hole accretion or compact star-formation activity remains a topic of intense debate.  Specifically, some studies indicate that obscured AGN dominate the LRD population \citep{Kocevski2024}, while others suggest that many/most LRDs are dominated by starlight (motivated in part by the fact that the SED inflection typically occurs at $\lambda_{\rm rest} \simeq 4000$\AA\ as expected from a Balmer break \citep{Setton2024}, and by the lack of evidence for hot AGN-heated dust at longer (MIRI) wavelengths; \citealt{PerezGonzalez2024b}). Although LRDs are rare enough that their integrated contribution to overall dust-obscured cosmic SFRD is considered to be small/negligible \citep{Casey2024}, their impact on cosmic SFRD becomes potentially more significant at very high redshifts ($z > 5$; \citealt{Williams2024}). Hence, the existence/prevalence of LRDs in red colour selections of high-redshift objects must be considered carefully.

Finally, we note that, ideally, the contribution of dust-enshrouded SFRD would be based on complete and deep surveys of dust {\it emission} (i.e., from far-infrared/mm observations) rather than dust {\it attenuation} (as is the case in the colour selection of red galaxies). However, the progress in completing our understanding of the prevalence of dust emission at very high redshifts remains limited. The existence of bright sub-mm galaxies has been known since the advent of SCUBA \citep{Holland1999} on the James Clerk Maxwell Telescope (JCMT) in the late 1990s \citep{Smail1997, Hughes1998}, and, helped by the subsequent advent of SCUBA-2 \citep{Holland2003}, the prevalence of such (sub)mm-bright objects is now well established \citep{coppin2007, Geach2017, Michalowski2017, Simpson2019, Simpson2020}. However, such rare, extreme objects only scratch the surface of the full contribution of dust-enshrouded star formation activity to cosmic SFRD, as has been revealed by deeper (sub)mm surveys with ALMA \citep{Dunlop2017, Bouwens2020}. These ALMA studies have enabled the direct detection of dust-enshrouded star-forming galaxies an order-of-magnitude fainter than the bright (sub)mm sources uncovered by the aforementioned single-dish surveys, reaching down to star-formation rates SFR $\simeq 40\,{\rm M_{\odot}\,yr^{-1}}$, with stacking in stellar mass bins enabling even deeper statistical detections \citep{McLure2018}. However, deep, contiguous ALMA surveys remain very limited in area (and hence sample relatively small cosmological volumes), while the pointed ALMA follow-up of UV-selected galaxies (e.g., the ALPINE or REBELS programmes: \citealt{Gruppioni2020}; \citealt{Bouwens2022}) is incomplete and is explicitly biased against the observation of the reddest dust-obscured sources. One consequence of this is that, as mentioned above, while the inventory of dust-enshrouded star-formation activity around cosmic noon is now reasonably well established \citep[e.g.,][]{Dunlop2017, Zavala2021}, the situation at $z > 3$ has remained much less clear. 

In this study, we select red galaxies using the eight-band {\it JWST} PRIMER NIRCam imaging (Dunlop et al., in prep.) based on JWST-derived colour criteria. Section \ref{observations} summarises the {\it JWST} and {\it HST} imaging datasets used in this study, along with the process of multi-wavelength catalogue production and subsequent galaxy photometric redshift estimation. In Section \ref{physicalproperties}, we then derive and discuss the physical properties of red galaxies in the context of the overall PRIMER-selected galaxy sample. Next, in Section \ref{Secnumber_density}, we explore potential evolutionary connections between dusty red objects and lower-redshift quiescent galaxies by comparing their comoving number densities. Lastly, in Section \ref{Sec_SFRD} we estimate the contribution of red galaxies to the SFRD and compare with the total SFRD also computed in this study. Our conclusions are summarised in \ref{Summary}. 

Throughout the paper, we assume a flat cold dark matter cosmology with $H_0 = 67.4\,{\rm km\,s^{-1}\,Mpc^{-1}}$, ${\rm \Omega_m} = 0.315$ and ${\rm \Omega_{\Lambda}} = 0.685$
(Planck Collaboration et al. 2020). All quoted magnitudes are in the AB system \citep{Oke83}, and all derived star-formation rates (SFR) and stellar masses ($M_*$) assume a \citet{Kroupa2001} IMF.

\section{Data: the COSMOS PRIMER survey}
\label{observations}

\subsection{Source extraction, photometry and photometric redshifts}

The galaxy catalogue for use in this study was extracted from the {\it JWST} 8-band NIRCam imaging of the central 142\,arcmin$^2$ of the COSMOS survey field provided by PRIMER (Dunlop et al. in prep), along with the 3-band optical imaging of the same area originally delivered by the {\it HST} CANDELS Treasury programme \citep{Grogin11, Koekemoer2011}. Our source catalogue is thus based on 11-band photometry from the {\it JWST} NIRCam F090W, F115W, F150W, F200W, F277W, F356W, F410M, and F444W imaging, and the {\it HST} ACS F435W, F606W, F814W imaging (after PSF homogenisation to the F444W image resolution). 

The PRIMER NIRCam data were reduced using the PRIMER Enhanced NIRCam Image Processing Library ({\sc pencil}; Magee et al., in preparation, Dunlop et al. in preparation) software. The astrometry of all the reduced images was aligned to {\it Gaia} Data Release 3 (Gaia Collaboration et al. 2023) and stacked to the same pixel scale of 0.03\,arcsec.

The NIRCam source selection utilised here is based on the F356W imaging. At F356W, the NIRCam detection has a sensitivity of $\sigma=  0.0039\,{\rm \mu}$Jy (point-source total, based on 0.5-arcsec diameter apertures) and an angular resolution of $0.12$\,arcsec. The 5-$\sigma$ point-source corrected depths of the {\it JWST} and {\it HST} photometry used to create the PRIMER COSMOS source catalogue used here are summarised in Table \ref{tab:depths}.

\begin{table}
\centering
\caption{The median global 5$\sigma$ limiting magnitudes for each of the {\it JWST} and {\it HST} bands used in this study. For all \textit{JWST} NIRCam and \textit{HST} imaging, these are measured through $0.5^{\prime\prime}-$diameter apertures and then corrected to total assuming a point-source correction. 
}
\label{tab:depths}
    \begin{tabular}{lc}
    \hline
    Filter & 5$\sigma$ limit (AB mag) \\
        \hline
    HST ACS F435W & 27.4\\
    HST ACS F606W & 27.5\\
    HST ACS F814W & 27.3\\
    JWST NIRCam F090W & 27.4\\
    JWST NIRCam F115W & 27.6\\
    JWST NIRCam F150W & 27.8\\
    JWST NIRCam F200W & 27.9\\
    JWST NIRCam F277W & 28.1\\
    JWST NIRCam F356W & 28.2\\
    JWST NIRCam F410M & 27.5\\
    JWST NIRCam F444W & 27.9\\
    \hline
    \end{tabular}
\end{table}

The process of catalogue construction is described in detail by \citet{Begley2025}. In brief, following PSF homogenisation to the angular resolution of the F444W imaging, we constructed a multi-frequency catalogue by running \textsc{Source Extractor} in dual-image mode (using the F356W data as the detection image) and extracted aperture photometry through 0.5-arcsec diameter apertures. To account for light outside the aperture for extended sources, we require a further correction to the total beyond point-source corrections. Following \citep{McLeod2024}, we scaled the fluxes (and therefore also SED-derived properties such as the stellar mass) to the FLUX\_AUTO value \citep{Kron1980} measured by \textsc{Source Extractor}. We added a further 10\% to account for light beyond the Kron aperture (see \citet{McLeod2024} for details). To minimise the number of spurious sources in the catalogue, we retained only those objects detected with S/N $>3$ in the F356W band. The final PRIMER COSMOS NIRCam-selected galaxy sample used here contains $\mathrm{ \sim 43,000}$ sources. 

Finally, photometric redshifts for all galaxies in the sample were estimated using \textsc{EAZY-PY} \citep{Brammer08} with the combined 11-band {\it JWST}+{\it HST} photometry. Specifically, the assigned photometric redshift for each source is computed as the uncertainty-weighted combination of the best-fitting redshifts from four different \textsc{EAZY-PY} configurations (including templates from \citet{Larson2023} and \citet{Hainline2024}; see \citet{Begley2025} for further details).

\subsection{Red galaxy selection}
\label{redgalaxies_selection}

\begin{figure}
 \centering
     \includegraphics[width=\columnwidth]{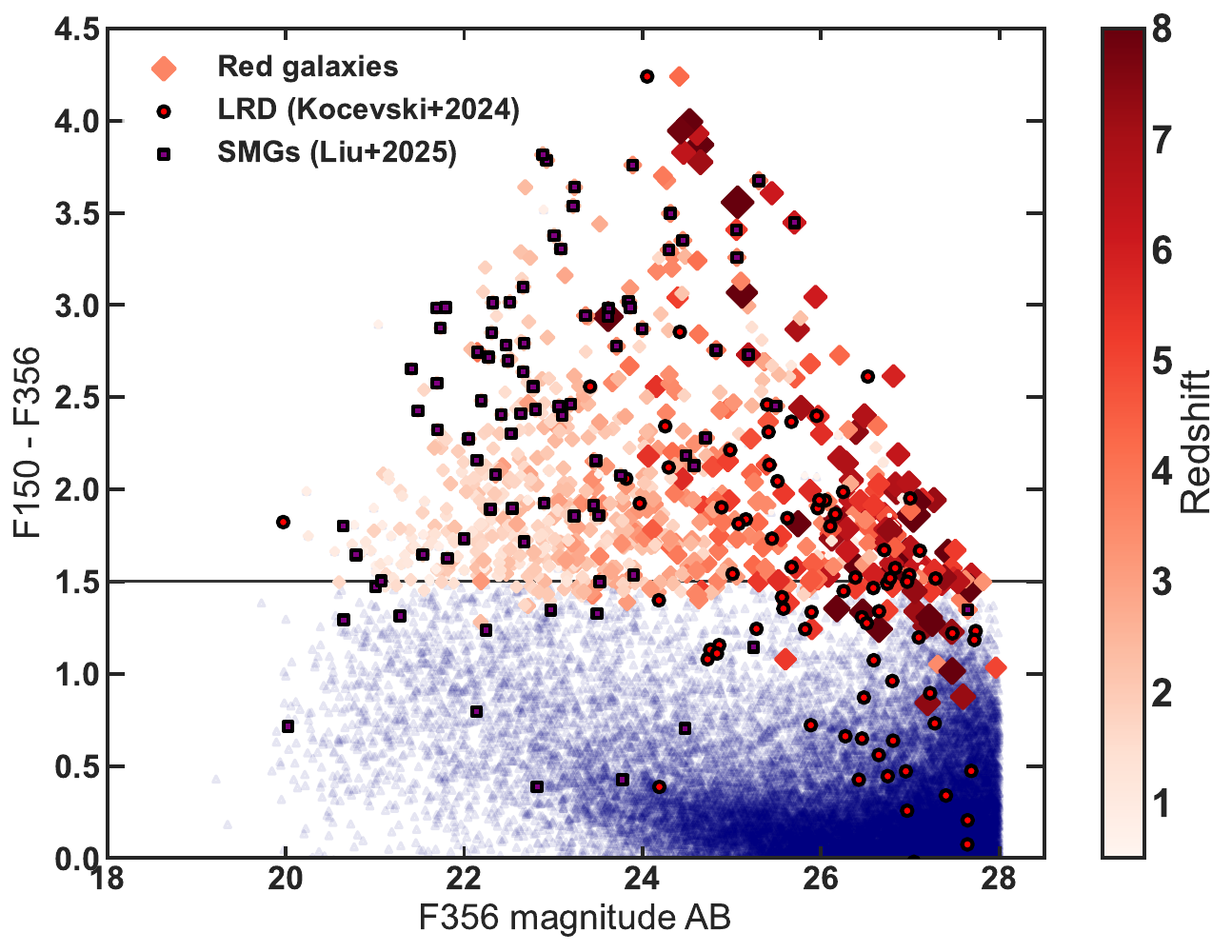 } \caption{Colour–magnitude diagram of the full PRIMER (COSMOS) sample, showing F356W magnitude versus F150W–F356W colour. Blue triangles represent all sources, while red diamonds highlight the red colour-selected subsample. The colour scale indicates photometric redshift. The black line marks the $\mathrm{F150W - F356W > 1.5}$ selection criterion from \citet{PerezGonzalez2023}, which identifies the majority of red galaxies but misses some heavily dust-obscured, high-redshift sources with strong F444W emission. These are recovered using an extended criterion based on $\mathrm{F150W-F444W> 2}$ \citep[e.g.,][]{Barrufet2023, Gottumukkala2024, Williams2024}. Unlike dropout-based selections, this approach includes all red galaxies irrespective of flux at $\mathrm{1.5\,\mu m}$, allowing for a more complete census of red populations across cosmic time. Purple squares denote submillimeter galaxies (SMGs) from \citet{Liu2025}, 82 \% of which satisfy the red selection. Red circles correspond to Little Red Dots (LRDs) from \citet{Kocevski2024}, with approximately half meeting the red criterion. 
 }  
   \label{Figcolourselection}
\end{figure}

The first two years of {\it JWST} data have revealed many red sources, potentially indicative of dust content missed by previous surveys. In this section, we apply several red colour criteria from the literature to maximise the inclusion of red sources. Figure \ref{Figcolourselection} shows the total sample of 777 red sources, primarily identified using colour selection criteria from \citet{PerezGonzalez2023, Barrufet2023, Gottumukkala2024, Williams2023}. The selection $\mathrm{F150W-F356W > 1.5}$ identifies the majority of red sources (702) but excludes certain high-redshift dusty sources captured by $\mathrm{F150W-F444W > 1.75}$; this is a similar criterion to that adopted by  \citet{Barrufet2023, Gottumukkala2024, Williams2023} and adds 77 red sources. Importantly, no cut-off in shorter wavelength colours was applied (as used to select HST-dark galaxies), including red sources across all redshifts. This comprehensive selection allows for a detailed study of the evolution of red sources over cosmic time. 

Cross-matching our red source catalogue with the SMG catalogue from \citet{Liu2025}, we find that 77 \% of SMGs meet our red colour selection criteria. However, as the \citet{Liu2025} sample includes both bright submillimeter sources and fainter DSFGs, this fraction likely represents an upper limit. Notably, only 16 \% of SMGs with $\mathrm{S_{870,\mu m} > 3.5,mJy}$ are selected as red. Unsurprisingly, most SMGs occupy the left side of the colour diagram, corresponding to more massive galaxies and lower to moderate redshifts (see Figure \ref{Figcolourselection}). 
Furthermore, we find no overlap between SMGs and LRDs, as expected given the lack of direct dust emission detected from LRDs at far-infrared or submillimeter wavelengths to date. 

We cross-matched the catalogue of 81 LRDs in the COSMOS field from \citet{Kocevski2024} with our red source catalogue to identify which LRDs meet our colour selection (no extra sources were incorporated). We find that 46\% of the LRDs satisfy our colour selection criteria, while the remaining sources display significantly bluer colours (see Figure~\ref{Figcolourselection}). \citet{Kocevski2024} employed a more refined selection method based on the UV continuum slope ($\beta$), rather than broader colour-based criteria such as the V-shape selection used here. However, their COSMOS field sample consists primarily of photometric redshifts (only one source has a spectroscopic redshift). We also evaluated the LRD population using the selection criteria from \citet{Greene2023} and \citet{Williams2024}, identifying 28 additional LRDs, all of which meet our red-source criteria. Approximately half were identified by each study, with only one overlapping source. This overlap highlights the need for more standardised and comprehensive selection criteria to capture the full range of the LRD population.

\subsection{Spectroscopic redshifts}
\label{Section_zspec}

\begin{figure}
 \centering
     \includegraphics[width=\columnwidth]{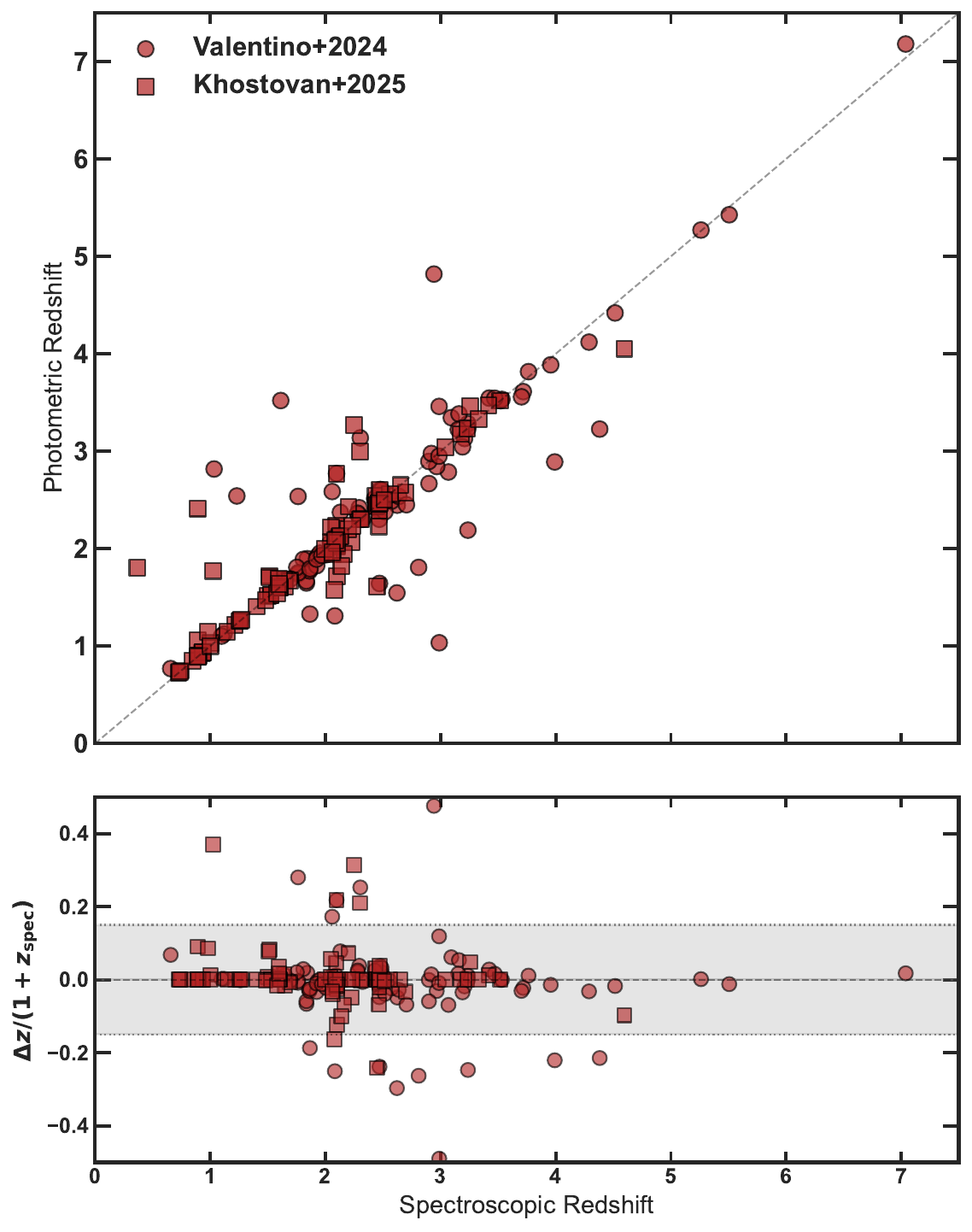} \caption{Comparison between spectroscopic and photometric redshifts for 193 red galaxies in the PRIMER COSMOS field. Photometric redshifts, computed as the combination of four independent estimates, are plotted against spectroscopic redshifts from the DAWN JWST Archive (DJA; red circles, 93 sources) and the pre-JWST compilation of \citet{Khostovan2025} (red squares, 100 sources). The one-to-one relation is shown as a dashed black line. The lower panel shows the residuals $\Delta z / (1+z_{\mathrm{spec}})$, with horizontal dotted lines marking the outlier boundary at $\pm 0.15$. 
 }  
   \label{zspec_zphot}
\end{figure}

To assess the accuracy of our photometric redshifts, Figure \ref{zspec_zphot} compares them with the spectroscopic redshifts for $\simeq 25$\% of the red galaxy sample. In detail, spectroscopic counterparts for our photometric sources in the COSMOS field were identified using the reduced \textsc{NIRSpec/PRISM} spectra available in the \textsc{DAWN JWST} Archive \citep[DJA][]{Heintz2025}. We utilized version-3 spectra\footnote{\url{https://s3.amazonaws.com/msaexp-nirspec/extractions/nirspec_graded_v3.html}}, uniformly processed with \textsc{MSAEXP} software; for details, see \citet{Heintz2025}, and \citet{Valentino2025}. We selected \texttt{grade = 3} \textsc{PRISM} spectra from the DJA, which have undergone visual inspection and have been confirmed to provide unambiguous redshifts. A one-arcsec matching radius was applied to identify spectroscopic counterparts, yielding 16 matches to \textsc{PRISM} spectra from the following JWST programs: \textsc{PID GO-2565} (PI: Glazebrook), \textsc{PID GTO-1214} (PI: Luetzgendorf), and \textsc{PID DD-6585} (PI: Coulter). We identified 93 red sources with high-confidence spectroscopic redshifts from the DJA archive. Additionally, we incorporated the \citet{Khostovan2025} compilation, a pre-JWST spectroscopic redshift database for the COSMOS field, aggregating data from 108 programs up to $\mathrm{z \sim 8}$, yielding 93 red galaxies with spectroscopic redshifts at a 97 \% confidence level. There are 14 sources repeated in the DJA catalogue, all in good agreement with the spectroscopic redshift. We evaluated the accuracy of our photometric redshift using this combined dataset of 187 spectroscopic redshifts from \citet{Khostovan2025} and the DJA archive. The agreement between the two redshift estimates is reflected in a robust normalised scatter of $\sigma_{\Delta z / (1+z)} = 0.02$, and an outlier fraction of $13\%$, using the standard threshold of $|\Delta z| / (1+z) > 0.15$ (see Figure \ref{zspec_zphot}). Note that estimating photometric redshifts for red sources is inherently more challenging due to the shape of their SEDs. As a result, it is not uncommon to observe a higher fraction of outliers, especially when the sample includes LRDs.

We applied the same quality criteria for the non-red sources in our catalogue. Using a one-arcsecond matching radius, we identified $\sim 1600$ spectroscopic counterparts from \citet{Khostovan2025} and $\sim 980$ from \citet{Valentino2025}. After excluding $\sim 120$ duplicate matches, we obtained a total of $\sim 2500$ unique sources with high-confidence spectroscopic redshifts. The spectroscopic redshifts account for $\sim 6 \%$ of the total sample of galaxies.

\subsection{Spectral energy distribution fitting}
\label{SED_fitting}

We employed a general SED fitting approach suitable for modelling the diverse galaxy population in the COSMOS field. The adopted star formation history (SFH) is a delayed model with an e-folding time $\tau$ ranging from 0.1 to 9 Gyr, enabling representation of both short bursts and nearly constant SFHs. Stellar population models are based on the updated \citet{Bruzual2003} library with a \citet{Kroupa2001} initial mass function (IMF). We considered metallicities ranging from 0.2 to 2.5 $\mathrm{Z_{\odot}}$. Nebular continuum and emission lines were included using the CLOUDY photoionisation code \citep{Ferland2017}, with an ionisation parameter set to $\log U = -2$. Dust attenuation was modelled with the \citet{Calzetti2000} law, allowing $\mathrm{A_V}$ to range from 0 to 6 mag, thereby encompassing both dusty and minimally attenuated galaxies. 

The redshifts used in our SED fitting procedure were based on the photometric redshifts described in Section~\ref{Section_zspec}. These were allowed to vary within their respective uncertainties during the fitting process, with the final adopted redshift calculated as the weighted average of the allowed range. In contrast, spectroscopic redshifts (Section~\ref{Section_zspec}) were fixed to their measured values with no uncertainty, thus precluding variation during the fitting. This approach represents a balance between leaving the redshift completely free and fixing it precisely, enabling a realistic assessment of redshift uncertainties while ensuring reliable fitting results. 

After performing the SED fitting, we applied quality control criteria to retain only reliable fits. Specifically, we removed cases of poorly constrained SED fits, including red sources incorrectly fitted to very high redshifts ($z>10$), a known degeneracy for certain red galaxies at $z\sim4$–$5$. Additionally, we excluded sources exhibiting physically implausible SED fits. Overall, our analysis is not focused on extreme individual cases but rather aims to characterise the broader properties of the galaxy population robustly.

\newpage
\section{Physical properties of red galaxies. } 
\label{physicalproperties}

This section assesses the properties of red galaxies in comparison to those from the full PRIMER sample. Both red and total galaxy properties are derived using a uniform methodology (see Section \ref{SED_fitting}), ensuring a rigorous and consistent comparison. This approach evaluates the effectiveness of colour selection criteria and identifies dusty, massive galaxies that might not conform to traditional red colour selections. 

\begin{figure}
 \centering
     \includegraphics[width=\columnwidth]{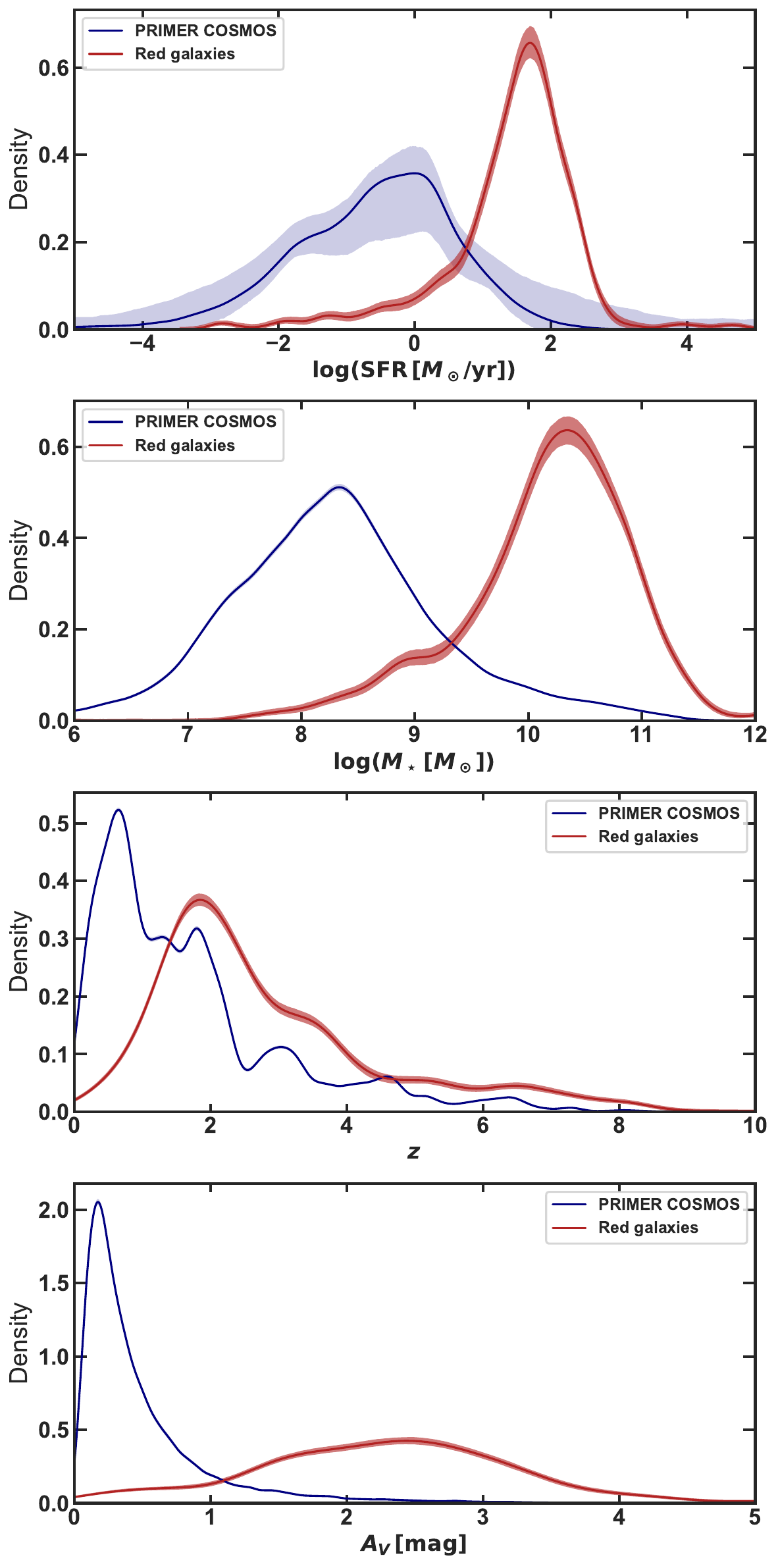}    \caption{Kernel density estimates showing the distributions of sources in the PRIMER COSMOS field, comparing the full sample (blue) and the red galaxy subsample (red) as a function of SFR, stellar mass, photometric redshift and dust attenuation. Red galaxies exhibit $\mathrm{SFR \sim 40 M_{\odot}yr^{-1}}$ and stellar masses of $\mathrm{\log(M_{*}/M_{\odot})\sim 10.3}$, both exceeding the medians of the full JWST-detected population by more than two orders of magnitude. Their median redshift is $\mathrm{z\simeq2.3}$, consistent with the peak of cosmic star formation ("cosmic noon"), compared to $\mathrm{z\sim1.3}$ for the overall sample. 
 These properties confirm that the red population shares the typical characteristics of dusty star-forming galaxies, with significant mass, obscuration, and ongoing star formation activity.} 
   \label{Figkdeplot}
\end{figure}

Figure \ref{Figkdeplot} presents kernel density estimate (KDE) plots comparing the distributions of SFR, stellar mass, redshift and dust attenuation ($\mathrm{A_{v}}$) for red galaxies and the overall COSMOS sample. Confidence intervals are derived from Monte Carlo (MC) sampling of individual galaxy measurements. The distribution of each inferred parameter is MC sampled, and the KDE recalculated for each MC realisation, repeating this process $10^4$ times to ensure robust estimates and accurate representation of uncertainties. The red galaxy subset shows significantly higher $\mathrm{SFR \sim 40 M_{\odot}/yr}$, nearly two orders of magnitude larger than the COSMOS median of $\mathrm{SFR \sim 0.2  M_{\odot}/yr}$. Notably, red galaxies have a median stellar mass of $\mathrm{log(M_{*}/M_{\odot}) = 10.3^{+0.6}_{-0.8}}$, significantly exceeding the general sample median of $\mathrm{log(M_{*}/M_{\odot}) = 8.3^{+0.9}_{-0.9}}$, demonstrating that that the red colour selection effectively identifies massive galaxies.  However, this selection does not capture all massive galaxies as seen by the tail of the total galaxy distribution extending above $\mathrm{log(M_{*}/M_{\odot}) > 10}$. The median redshift for red galaxies is $\mathrm{z = 2.3^{+2.2}_{-0.8}}$, higher but statistically consistent with the overall galaxy population ($\mathrm{z = 1.3^{+1.7}_{-0.8}}$) due to overlapping uncertainties. This differs from previous JWST studies that primarily identified red galaxies at $\mathrm{z > 3}$ \citep{Barrufet2023}. The discrepancy likely arises from differences in selection criteria: earlier studies focused specifically on HST-dark (or faint) galaxies, imposing magnitude cuts fainter than $\sim26-27$ mag in F160W (observed at $\mathrm{1.6~\mu m}$), whereas our sample does not apply a magnitude restriction. Importantly, the dust attenuation in red galaxies is significantly larger with $\mathrm{A_{v} = 2.3 ^{+0.9} _{-0.9}}$ compared to $\mathrm{A_{v} = 0.3 ^{+0.5} _{-0.2}}$ for the general sample, indicating substantial dust content in red galaxies. Nevertheless, approximately 9\% of red galaxies show \(\mathrm{A_{v} < 1~mag}\), indicating potential contamination by dust-poor quiescent galaxies. This fraction closely matches the spectroscopically confirmed quiescent galaxy fraction reported by \citet{Barrufet2024}, supporting the interpretation that a minority of red sources are truly quiescent rather than dusty star-forming. In the absence of spectra for the full sample, these galaxies cannot be reliably identified and are retained in the analysis, with the caveat that quiescent contamination is likely limited to \(\lesssim 10\%\).  

These results confirm that the red galaxy population identified in the COSMOS field represents a distinct subset of massive, dusty, star-forming systems near cosmic noon, effectively captured by the colour selection despite minor quiescent contamination.

\begin{figure}
 \centering
     \includegraphics[width=\columnwidth]{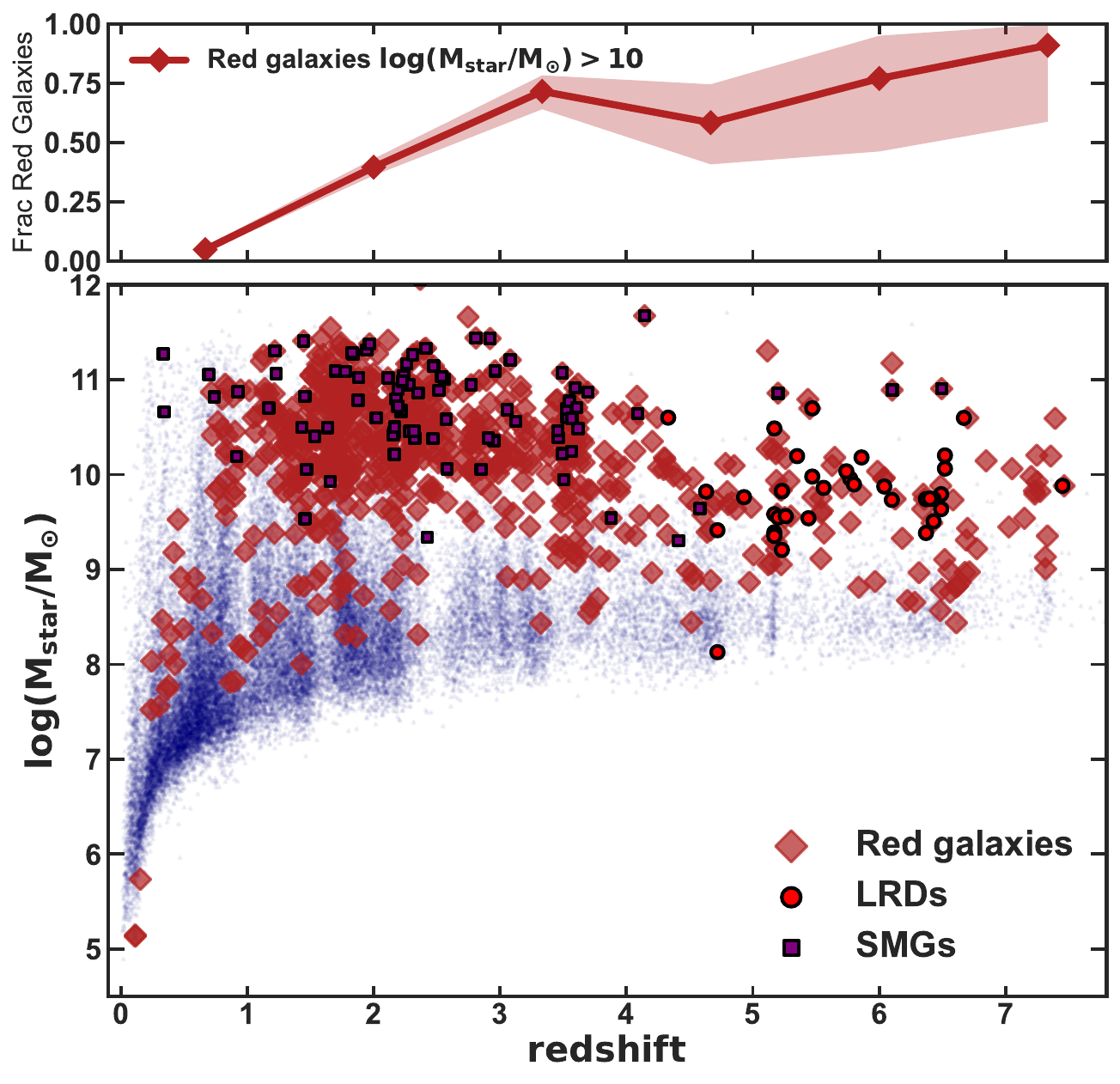}
 \caption{Stellar mass versus redshift for the red galaxies (red diamonds), SMGs (purple squares) and LRDs (red circles) and PRIMER COSMOS sample (blue triangles). Red galaxies dominate the high-mass region across all redshifts. SMGs are primarily found within this population at $\mathrm{z < 3}$, while LRDs contribute significantly at $\mathrm{z \sim 5-7}$. The top panel shows the fraction of red galaxies within the total sample as a function of redshift. The solid red line indicates the fraction of red galaxies among all galaxies with $\mathrm{log(M_{*}/M_{\odot}) > 10}$, demonstrating their clear dominance in the high-mass end at $\mathrm{z \sim 3}$. The red shaded area denotes binomial confidence intervals, which widen at $\mathrm{z > 3}$ due to limited source counts. Consequently, the fraction of massive red galaxies shows a robust increase up to $\mathrm{z \sim 3}$ is robust, while their evolution at higher redshifts remains less well constrained.}
   \label{Mstar_z}
\end{figure}

\begin{figure*}
 \centering
     \includegraphics[width=\textwidth]{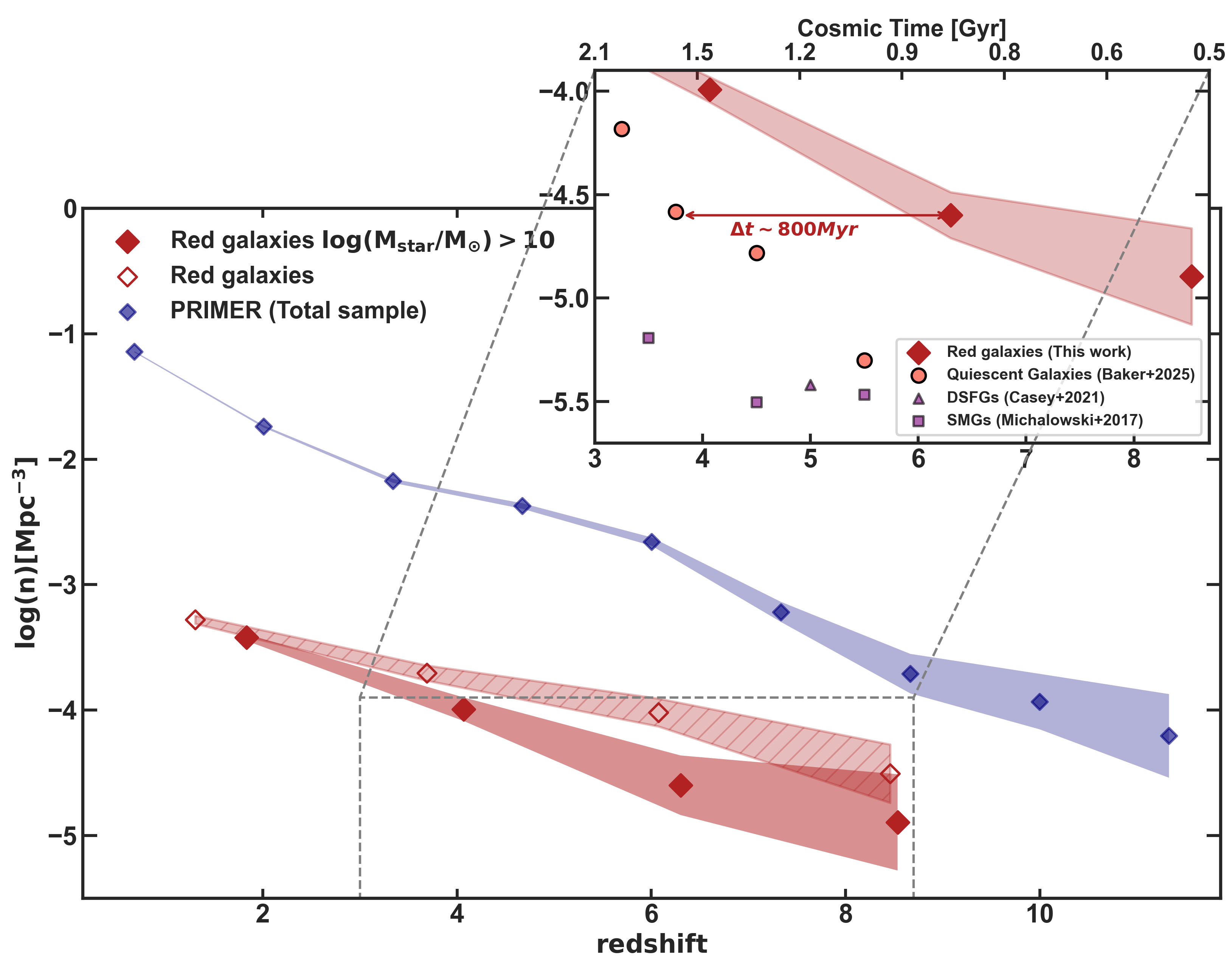}   
     \caption{Number densities as a function of redshift for several galaxy populations and surveys. Main Figure: The blue diamonds represent the number density of the total PRIMER-COSMOS sample, showing the expected cosmic decline of galaxies with increasing redshift. Red diamonds (filled: $\mathrm{\log(M_{*}/M_{\odot}) > 10}$; empty: all red galaxies) trace a different evolution, with a more moderate decrease and a broad peak around $\mathrm{z \sim 2}$. Shaded areas indicate uncertainties. Inset Figure: Zoomed-in view of the number density evolution at $\mathrm{z \gtrsim 3}$, comparing red galaxies from this work (red diamonds) with quiescent galaxies from \citet{Baker2025} (salmon circles; $\mathrm{\log(M_{*}/M_{\odot}) > 10}$), dusty star-forming galaxies (DSFGs; purple triangles) from \citet{Casey2019}, and submillimeter galaxies (SMGs; purple squares) from \citet{Michalowski2017}. At $\mathrm{z \sim 5}$, red galaxies are approximately one order of magnitude more numerous than DSFGs and SMGs. By $\mathrm{z \sim 6}$, their number density reaches $\mathrm{log(n) \sim -4~[Mpc^{-3}]}$, comparable to that of quiescent galaxies at $\mathrm{z \sim 4}$. The horizontal arrow indicates a time delay of $\Delta t \sim 800$ Myr between the two populations, suggesting that at least the abundance of red galaxies at high redshift is sufficient for them to evolve into the observed quiescent population.}
   \label{Fignumber_density}
\end{figure*}

\subsection{Red Galaxies Dominate the High-Stellar-Mass Regime}

We investigate the stellar mass-redshift distribution of red galaxies and the total galaxy population, highlighting the overlap with SMGs and LRDs. Furthermore, we quantify the fraction of massive galaxies as a function of redshift to evaluate how the colour-selected red galaxies trace the high-mass end of the galaxy distribution. 

Figure \ref{Mstar_z} presents the stellar mass-redshift distribution of the total galaxy sample, emphasising the red galaxies and their fractional contribution relative to the total sample. To evaluate the efficiency and evolution with redshift of the red colour selection in capturing massive galaxies, we divided both the total sample and the red sources at $\mathrm{log(M_{*}/M_{\odot}) > 10}$, above which the sample is expected to be complete across the redshift range considered. Uncertainties on the fractions are computed following \citet{Gehrels86} assuming the conservative case of maximal binomial variance. Red galaxies consistently occupy the high-mass regime across all redshifts, reflecting systematically greater stellar masses than the overall population. Our analysis shows that beyond $\mathrm{z > 3}$, red galaxies dominate the massive galaxy population, comprising 71 \% of massive galaxies at $\mathrm{z = 3.3}$ and increasing to 91 \% at $\mathrm{z \sim 7}$. However, we note significant uncertainties at $\mathrm{z > 3}$, in particular in the highest redshift bin, due to the limited number of sources ($\sim 30$). Therefore, we can only confidently assert a steep increase in the fraction of massive red galaxies over redshift up to $\mathrm{z \sim 3}$, with a likely more moderate increase at higher redshifts.

Cross-matching the red source sample with SMGs from \citep{Liu2025} and LRDs from \citet{Kocevski2024} (see Section \ref{redgalaxies_selection}) reveals a distinct stellar mass-redshift relation. SMGs are predominantly massive galaxies ($\mathrm{log(M_{*}/M_{\odot}) > 10}$) but are largely concentrated at $\mathrm{z < 3}$, consistent with findings from previous studies \citep[e.g.,][]{Dunlop2017}. In contrast, LRDs dominate at higher redshifts ($\mathrm{z \sim 5-7}$) while also appearing in the high-mass region of the diagram. This distribution likely reflects these populations' differing evolutionary stages and formation histories. As a caveat, stellar masses are uniformly derived in this study without accounting for possible AGN contamination, which may lead to a slight overestimation of LRD masses at $\mathrm{z > 5}$. 

While current colour selections recover most of massive galaxies, future studies should prioritise direct selection by physical properties —such as stellar mass and redshift— to achieve a more complete and unbiased census of the massive galaxy population.

\section{Could red dusty galaxies potentially be progenitors of quiescent galaxies at $\mathrm{\mathbf{\MakeLowercase{z} \sim 3}}$?}
\label{Secnumber_density}

\begin{figure*}
 \centering
     \includegraphics[width=0.8\textwidth]{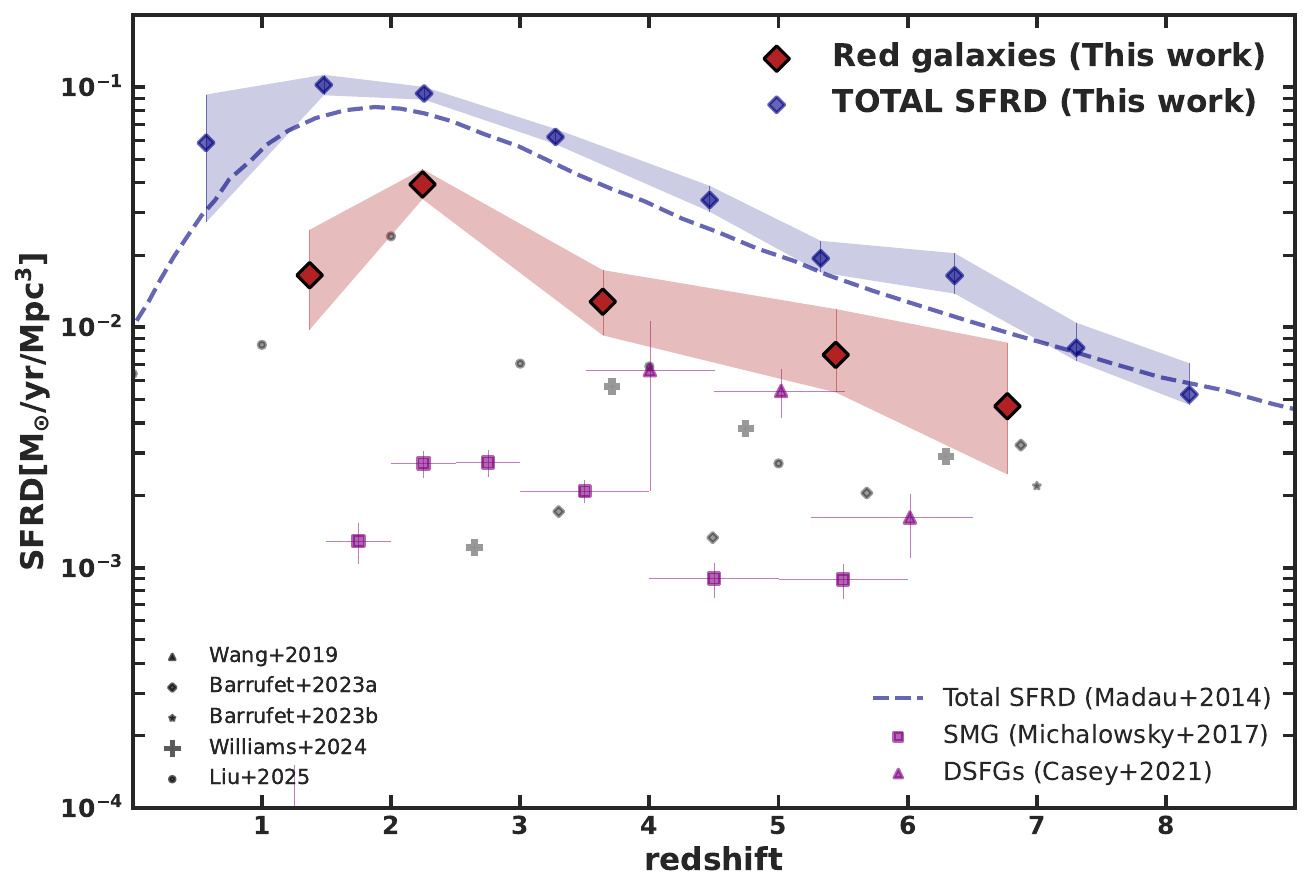}    
 \caption{Star formation rate density (SFRD) as a function of redshift. Blue diamonds represent the total SFRD for galaxies from the COSMOS-PRIMER field (this work), with shaded regions showing bootstrap-derived confidence intervals. This trend closely follows the total SFRD from \citet{Madau2014} (blue dashed line). Red galaxies (red diamonds) contribute significantly to the SFRD, accounting for approximately 40 \% at $\mathrm{z \sim 5}$. At $\mathrm{z \sim 5}$, the SFRD of red galaxies is comparable to that of DSFGs \citep[purple triangles;][]{Casey2021}, within the DSFGs' large uncertainties, and exceeds that of SMGs \citep[purple squares;][]{Michalowski2017} by an order of magnitude highlighting the significant contribution of red galaxies to the total SFRD. Despite lower individual SFRs, their high number density makes red galaxies a dominant contributor to the obscured SFRD across cosmic time.}
   \label{FigSFRD}
\end{figure*}

Over the past decade, independent studies have reported the emergence of both dusty \citep{Wang2016, AlcaldePampliega2019, Barrufet2023b} and quiescent galaxies  \citep{Merlin2019, Santini2019, Valentino2020, Carnall2020}. The advent of JWST has accelerated progress in both areas, with recent work confirming a significant population of dusty galaxies \citep{Barrufet2023b, Barro2023, PerezGonzalez2023, Nelson2022, Rodighiero2023, Gottumukkala2024, Akins2023, Barrufet2024} and spectroscopically confirming the existence of quiescent galaxies at $\mathrm{z >4}$ \citep{Carnall2023, Barrufet2024}. These findings demonstrate that both populations are more abundant than anticipated from pre-JWST surveys. However, the evolutionary connection between these massive red galaxies, both dusty star-forming and quiescent, remains uncertain within the broader context of high-mass galaxy formation.

NIRCam/JWST studies have explored the internal structures of a small sample of red galaxies \citep{Nelson2022, GomezGuijarro2023, PerezGonzalez2023, Sun2024}. These analyses suggest that some massive, dusty galaxies could plausibly evolve into quiescent galaxies at high redshift, although their spatially resolved properties and evolutionary timescales remain uncertain
Furthermore, pre-JWST work suggested extreme dusty galaxies, such as SMGs, as progenitors of quiescent galaxies at $\mathrm{z \sim 2}$ \citep{Toft2014, Valentino2020}. Building on this framework, we investigate whether more numerous, JWST-identified red dusty galaxies - characterised by lower SFRs than typical SMGs \citep{Barrufet2023} - may serve as viable progenitors of quiescent galaxies at $\mathrm{z \gtrsim 3-4}$. A critical requirement for this scenario is that their number density must match or exceed that of quiescent galaxies at $\mathrm{z > 3}$. 

To address the potential progenitor role of red galaxies, we derived the number densities of red galaxies, massive red galaxies ($\log(M_{\star}/M_{\odot}) > 10$), and the total galaxy population across cosmic time (see Fig. \ref{Fignumber_density}). The red galaxy sample was divided into four redshift bins ($z_{\rm med} = 1.3, 3.7, 6.1,$ and $8.5$), each containing more than 80 sources, except for the highest-redshift bin, which contains only 20 sources. Poisson statistics were used to estimate the uncertainty in the number density. The number density of red galaxies declines steadily with increasing redshift, from $\log(n/\mathrm{Mpc}^{-3}) = -3.28 \pm 0.03$ at $z \sim 1.3$ to $\log(n/\mathrm{Mpc}^{-3}) = -4.5 \pm 0.2$ at $z \sim 8.5$. This decline exceeds 5$\sigma$ significance, indicating a statistically robust evolutionary trend. Massive red galaxies exhibit a similar evolution, with $\log(n/\mathrm{Mpc}^{-3}) = -3.42 \pm 0.03$ at $z \sim 1.8$ and $-4.9 \pm 0.3$ at $z \sim 8.5$. Fig. \ref{Fignumber_density} shows that the total galaxy population follows similar redshift evolution up to $\mathrm{z < 6}$ but with number densities consistently higher, by $\mathrm{\sim 1.5 dex}$, than those of red galaxies. Interestingly, at $z \sim 6$, while the total galaxy number density drops more steeply at $z \sim 6$, the red galaxy population exhibits a smoother and more gradual decline. 

To assess whether massive red galaxies could plausibly evolve into quiescent galaxies at $z>3$, we compare our measured number densities with those of quiescent galaxies reported by \citet{Baker2025}. The inset of Fig. \ref{Fignumber_density} shows that the number density of massive red galaxies ($\log(n/\mathrm{Mpc}^{-3}) \approx -4.5$) at $z \sim 6$ is comparable to that of quiescent galaxies with $\log(M_{\star}/M_{\odot}) > 10$ in the redshift bin $\mathrm{z=3.5-4.0}$. The $\sim$800 Myr cosmic time interval between these epochs provides more than sufficient time for quenching, especially considering that simulations predict quenching timescales of only $\sim$200–400 Myr \citep{Weller2025}, significantly shorter than our conservative estimate. In addition to the number densities, the median star formation rates of the red galaxies in the redshift bin of $\mathrm{z \sim 5-6}$ are broadly consistent with those of simulated progenitors of high-redshift quiescent galaxies reported in \citep{Lagos2024b}. The slightly higher SFRs in some simulations likely reflect differences in star formation history assumptions. Despite these variations, the agreement supports the interpretation of dusty red galaxies as plausible precursors to quiescent galaxies at $z > 3$. 

For context, we also consider dusty star-forming galaxies (DSFGs) and SMGs at similar redshifts from \citet{Casey2021} and \citet{Michalowski2017}, respectively. These extreme dusty galaxies exhibit number densities below $\log(n/\mathrm{Mpc}^{-3}) \sim -5$ over $3 < z < 6$, significantly lower than those of quiescent galaxies at $z > 3$. This suggests that DSFGs and SMGs are unlikely to be the dominant progenitors of the high-redshift quiescent population, given both their limited abundance and the short available cosmic time for quenching. 

In summary, our results demonstrate that red massive galaxies have number densities at high redshift consistent with those required to explain the observed population of massive quiescent galaxies at $z > 3$. However, number density alone is not sufficient to establish a progenitor-descendant link. A deeper analysis of their physical properties, star formation histories and structural evolution will be necessary to confirm their evolutionary connection.

\section{Contribution of red galaxies to the total star formation rate density. } 
\label{Sec_SFRD}

Quantifying the SFRD of dusty galaxies across cosmic time is essential for understanding the global buildup of stellar mass in the Universe. The evolution of the cosmic SFRD follows a well-defined trend: a rapid rise from early epochs, peaking around $\mathrm{z \sim 2}$ (the so-called “cosmic noon”), followed by a steady decline towards the present day \citep{Madau2014}. While the contribution of dusty galaxies is traditionally traced via direct dust emission at far-infrared and submillimetre wavelengths, their role at $\mathrm{z > 3}$ remains uncertain. Early {\it JWST} studies have begun to probe the SFRD of red galaxies \citep{Barrufet2023, Williams2023}, suggesting they may contribute more significantly than previously estimated. Moreover, \citet{Williams2023} show that for galaxies with $\mathrm{SFR > 100 M_{\odot},yr^{-1}}$, inclusion of far-IR and submillimetre data increases SFR estimates by less than a factor of 10, while for galaxies with $\mathrm{SFR < 100 M_{\odot},yr^{-1}}$, the SFR remains unchanged whether or not these wavelengths are included. In this section, we compute the SFRD of red galaxies and compare it with that of the total galaxy population, applying a consistent methodology across redshift. While $\sim$90 \% of red galaxies in our sample are dusty star-forming galaxies, a small fraction are quiescent, and our selection misses $\mathrm{\sim 20\%}$ of the broader infrared-luminous population (see Section \ref{physicalproperties}). As such, the derived SFRD should be interpreted as a conservative lower limit on the total contribution from dusty galaxies.

In Figure~\ref{FigSFRD}, we present the total SFRD from the PRIMER–COSMOS field alongside the contribution from the 777 red galaxies. To quantify the contribution of red galaxies to the cosmic SFRD, we divide the red galaxy sample into five equidistant redshift bins spanning $z \sim 0.1$–8, each containing between $\mathrm{N} \sim 50$ and $\mathrm{N} \sim 340$ galaxies. We adopt a bootstrap resampling method to estimate uncertainties, drawing 1000 repetitions of the SFR and redshift distributions for each bin to account for sample variance. The resulting SFRD values for red galaxies, computed in the five redshift bins centred at $z = [1.37,\ 2.24,\ 3.64,\ 5.77,\ 6.77]$, are $\mathrm{SFRD} = [1.7^{+0.9}_{-0.7},\ 3.9^{+0.6}_{-0.5},\ 1.3^{+0.5}_{-0.4},\ 0.8^{+0.5}_{-0.2},\ 0.5^{+0.4}_{-0.2}] \times 10^{-2}\,M_\odot\,\mathrm{yr^{-1}\,Mpc^{-3}}$.

For comparison, the total PRIMER–COSMOS sample is divided into nine redshift bins using the same methodology and bootstrap resampling approach as for the red galaxy sample. The resulting total SFRD closely follows the canonical evolution reported by \citet{Madau2014}. This agreement confirms that our analysis captures the bulk of the star formation activity in the Universe, demonstrating both the completeness of the parent sample and the reliability of our method. This methodological consistency is crucial for accurately assessing the relative contributions of specific galaxy populations —such as red galaxies— to the total SFRD, particularly at early cosmic times, which is the central objective of this section. The red galaxy population contributes significantly to the cosmic SFRD across redshift, reaching a peak of $\mathrm{3.9 \times 10^{-2}\ M_{\odot},yr^{-1},Mpc^{-3}}$ at $\mathrm{z = 2.2}$. Notably, this coincides with the epoch when the cosmic SFRD is dominated by obscured star formation \citep[e.g.][]{Madau2014}, and our red galaxy SFRD nearly accounts for the total, underscoring their dominant role at cosmic noon. The slight shortfall may reflect the fact that our selection recovers only $\sim 80 \%$ of SMGs, which predominantly peak at $\mathrm{z \sim 2}$. Additionally, the $\mathrm{F150W - F356W > 1.5}$ colour selection is optimised for identifying red galaxies at $\mathrm{z > 3}$; at lower redshifts, the Balmer break falls blueward of F150W, potentially causing us to miss more typical dusty star-forming galaxies due to selection effects. At $\mathrm{z = 5.4}$, red galaxies contribute $\sim 40 \%$ of the total SFRD, and still account for $\sim 34 \%$ at $\mathrm{z = 6.8}$, despite the global decline in star formation. This persistent contribution out to $\mathrm{z \sim 7}$ highlights red galaxies as a significant component of the cosmic star formation budget at early times.

To evaluate the impact of LRDs on the cosmic SFRD, we computed the red galaxy contribution both including and excluding these sources (as identified by \citealt{Kocevski2023}. At $z \sim 5.4$, excluding LRDs lowers the SFRD from $\mathrm{7.7 \times 10^{-3}}$ to $\mathrm{4.8 \times 10^{-3}}~M_{\odot},\mathrm{yr^{-1},Mpc^{-3}}$, a $\sim38 \%$ reduction, though still within the uncertainty range. This contrast is less pronounced than that reported by \citet{Williams2024}, likely due to our nearly tenfold larger red galaxy sample, which mitigates statistical fluctuations. We also highlight that the majority of LRDs considered in this study have only photometric redshifts \citep[e.g., see][]{Kocevski2025}; larger numbers of spectroscopic redshift confirmations will enable sample improvements in future works. Interestingly, only 46\% of the LRDs from \citet{Kocevski2025} fall within our red galaxy selection, indicating that a significant fraction lies outside our colour and magnitude criteria. This is consistent with findings from \citet{PerezGonzalez2024b}, where recovering the full LRD population required reaching down to $\mathrm{F356W} \sim 27$ mag, and with the colour cut of $\mathrm{F150W-F356W} > 1$ mag adopted in \citet{PerezGonzalez2024a}, both less restrictive than the selection used here. In conclusion, while LRDs do not significantly alter the SFRD at high redshift in our large sample, their contribution can be non-negligible in smaller datasets, and it should be considered carefully, preferably supported by spectroscopic confirmation. 

We investigate the origin of discrepancies in the SFRD reported for various dusty galaxy populations. Comparing our red galaxy measurements with those of SMGs from \citet{Michalowski2017} and DSFGs from \citet{Casey2021}, we find that our SFRD values are consistently and significantly higher—by a factor of $\sim$4–10 at $\mathrm{z \sim 4}$. Despite substantial uncertainties in all datasets, particularly those relying on pre-JWST photometric redshifts, the observed excess cannot be explained by measurement scatter alone, indicating a genuine difference in the underlying galaxy populations or selection methods. These discrepancies may reflect the increasing depth and resolution of more recent surveys: SCUBA-2, used in earlier SMG studies, primarily detects bright, classical submillimeter galaxies, whereas ALMA probes fainter dusty sources, leading to improved completeness and higher SFRD estimates \citep[see][for a detailed review]{Casey2014}. This suggests that red sources might also be detected in dust emission with deeper observations, as \citet{Williams2024} demonstrated for a few sources. Our SFRD measurements lie consistently above those reported by \citet{Wang2019, Barrufet2023, Barrufet2023b, Williams2024, Liu2025} (grey points in Fig. \ref{FigSFRD}) across $3 \lesssim z \lesssim 6$, with differences from a factor of $\sim2$ up to nearly an order of magnitude. These differences arise from varying sample selections, survey depths, redshift estimation methods, and volume coverage. For instance, \citet{Wang2019} report lower values that are attributed to shallower data, smaller volumes, and larger photometric redshift uncertainties. \citet{Long2024} demonstrate that only $\sim40\%$ of DSFGs would be recovered using an HST-dark colour selection (i.e \citet{Wang2019} selection), highlighting the incompleteness of colour-limited samples. Our results suggest that red galaxies, as selected in this work, include dusty systems that may be missed by such strategies.

Among JWST-based studies, our SFRD values exceed those of \citet{Barrufet2023}, likely due to our larger sample size and more accurate photometric redshifts. Their smaller red galaxy sample is strongly affected by the inclusion of LRDs, whose presence can strongly influence high-redshift SFRD estimates at $\mathrm{z>5}$. Similarly, \citet{Williams2024} reports lower SFRDs at higher redshifts; their colour criteria specifically target red galaxies at $z>3$, thereby excluding lower-redshift dusty galaxies. This selection difference contributes to the nearly two orders of magnitude discrepancy at $z \sim 2$. In contrast, \citet{Barrufet2023b} reports a single SFRD point at $z \sim 7$ based on a UV-bright selected sample with ALMA detections. Although their value is similar to ours, the underlying galaxy populations are fundamentally different. This suggests that UV-bright and red, dusty galaxies may represent complementary subsets of the star-forming population at high redshift, and that combining both might be necessary to fully recover the SFRD at $z \sim 7$. Finally, the submillimeter-selected sample of \citet{Liu2025} yields lower SFRDs across redshift, as their selection targets only the brightest end of the dusty population with clear submillimeter detections. Thus, our red galaxy selection likely provides a more comprehensive assessment of dusty star formation, including sources missed by UV, colour, or submillimeter-based selections.
 
In summary, our analysis demonstrates that red galaxies make a substantial contribution to the cosmic SFRD over a broad redshift range. This significant contribution persists even when LRDs are excluded, highlighting clear population differences compared to other galaxy selections in the literature. Our results emphasise the importance of consistent, well-characterised samples to robustly trace obscured star formation and support the interpretation that red galaxies consistently play an important role in stellar mass assembly at high redshift.

\section{Summary and Conclusions} 
\label{Summary}

We have presented a comprehensive census and analysis of red galaxies in the COSMOS field, leveraging deep JWST PRIMER survey data combined with existing HST imaging. Our key results and their implications for galaxy evolution are:

\begin{itemize}
\item  Our colour selection identifies a robust sample of 777 red galaxies spanning $1 \lesssim z \lesssim 8$, with a median redshift of $\mathrm{z = 2.3^{+2.2}_{-0.8}}$, larger than the total COSMOS galaxy population. These red galaxies are massive, with a median stellar mass of $\mathrm{log(M_{*}/M_{\odot}) = 10.3^{+0.6}_{-0.8}}$, exceeding the population median by over two dex, and exhibit significantly higher dust attenuation with  $\mathrm{A_{v} = 2.3 ^{+0.9} _{-0.9}}$ mag. 

\item At fixed stellar mass, the dominance of red galaxies increases with redshift: they account for $72 \%$ of galaxies with $\log(M_*/M_\odot) > 10$ at $z  = 3.3$, rising to $ 91 \%$ by $z \sim 7$. Despite large uncertainties at high redshift due to small-number statistics, this trend demonstrates that massive, dusty star-forming galaxies dominate the high-mass end at early epochs.

\item The number density of massive red galaxies at $z = 6.1$, $\log n = -4.0 \pm 0.1 \mathrm{Mpc}^{-3}$, matches that of massive quiescent galaxies observed at $z \sim 4$, suggesting they are plausible direct progenitors of the $z > 3$ quiescent population. The cosmic time interval between these epochs,  $\sim 800\mathrm{Myr}$, is sufficient for typical quenching timescales, although matching number densities alone cannot confirm an evolutionary connection. 

\item Red galaxies make substantial contributions to the obscured cosmic star-formation rate density (SFRD), comprising up to $\sim 40 \%$ of the total SFRD at $z \sim 5$, with a peak at $z = 2.2$. 

\end{itemize}

Our results reveal that red galaxies are a major, previously under-represented population at high redshift, dominating the massive end and contributing significantly to the cosmic SFRD. Their fractional contribution to the total SFRD may peak at even higher redshifts, suggesting an increasingly critical role for red galaxies in stellar mass assembly during earlier cosmic epochs. These galaxies could represent a key transitional phase in the formation of massive quiescent galaxies. However, fully characterising their dust-obscured star formation and evolutionary paths will require deeper ALMA surveys, complementing the redshift and dust attenuation properties from JWST. Combining direct dust emission measurements from ALMA with JWST’s census of red galaxies will be key for building a complete picture of early galaxy evolution.

\section*{Acknowledgements}
LB, JSP and DJM acknowledge the support of the Royal Society
through the award of a Royal Society University Research Professorship to JSD. ACC acknowledge support from a UKRI Frontier Research Guarantee Grant (PI Carnall; grant reference EP/Y037065/1). RSE acknowledges financial support from the Peter and Patricia Gruber Foundation. FC, KZAC, DS and TS acknowledge support from a UKRI Frontier Research Guarantee Grant (PI Cullen; grant reference EP/X021025/1). RB acknowledges the support of the Science and Technology Facilities Council.  RKC is grateful for support from the Leverhulme Trust via the Leverhulme Early Career Fellowship.

Facilities: \textit{JWST}, \textit{HST}

Software:
    \texttt{matplotlib} \citep{matplotlib},
    \texttt{numpy} \citep{numpy},
    \texttt{scipy} \citep{scipy},
    \texttt{jupyter} \citep{jupyter},
    \texttt{Astropy}
    \citep{astropy1, astropy2},
    \texttt{grizli} \citep{grizli},
    \texttt{SExtractor} \citep{Bertin96},
    \texttt{bagpipes} \citep{Carnall2018}

\section*{Data availability}
All JWST and HST data products are available via the Mikulski
Archive for Space Telescopes (https://mast.stsci.edu). Additional data products are available from the authors upon reasonable request.


\bibliographystyle{mnras}
\bibliography{PRIMER_Redgalaxies} 

\bsp	
\label{lastpage}
\end{document}